# Ferroelectric and Magnetic Domains in $LuFe_2O_4$ Observed by Scanning Probe Microscopy

I. K. Yang[1], J. Kim[1,2], S. H. Lee[3], S. B. Kim[4], S.–W. Cheong[4,5], and Y. H. Jeong[1(a)]

[1]*Department of Physics, POSTECH - 77 Cheongam-Ro, Pohang 790-784, Republic of Korea*

[2]*CALDES, Institute of Basic Science, POSTECH - 77 Cheongam-Ro, Pohang 790-784, Republic of Korea*

[3]*YE team, Samsung Electronics - 1 Samsungjeonja-Ro, Hwaseong 445-330, Republic of Korea*

[4]*Laboratory of Pohang Emergent Materials, POSTECH - 77 Cheongam-Ro, Pohang 790-784, Republic of Korea*

[5]*Department of Physics and Astronomy, Rutgers University - Piscataway, New Jersey 08854, USA*

**Abstract** – $LuFe_2O_4$ is a multiferroic system which exhibits ferroelectricity, charge order, and ferrimagnetic order simultaneously below ~230 K. The ferroelectric domains of $LuFe_2O_4$ are imaged with both piezoresponse force microscopy (PFM) and electrostatic force microscopy (EFM), while the magnetic domains are characterized by magnetic force microscopy (MFM). Comparison of PFM and EFM results lead to a conclusion that the ferroelectricity is of electronic origin as opposed to the usual displacive one. Simultaneous characterization of ferroelectric and magnetic domains by EFM and MFM, respectively, on the same surface of $LuFe_2O_4$ reveals that both domains have irregular patterns of similar shape, but the length scales are quite different. The domain size is approximately 100 nm for the ferroelectric domains while the magnetic domain size is much larger and gets as large as 1μm. We also demonstrate that the origin of the formation of irregular domains in $LuFe_2O_4$ is not extrinsic but intrinsic.

[(a)]E-mail: `yhj@postech.ac.kr`

**Introduction.** – LuFe$_2$O$_4$ is a multiferroic compound which exhibits both ferroelectricity ($T_{FE}$ ≈ 340 K) and ferrimagnetism ($T_N$ ≈ 230 K) in a single phase at low temperatures [1]. This material has attracted much attention particularly because of its peculiar origin of ferroelectricity. Ferroelectric materials may be classified into proper and improper ferroelectrics depending on whether ferroelectricity itself is a primary order parameter or not. Many recently discovered multiferroics are improper ferroelectrics, and magnetic ordering is responsible for the ferroelectricity [2]. The ferroelectricity of LuFe$_2$O$_4$ also appears to be improper; it is claimed to be of electronic origin and caused by charge ordering (CO) [3]. This class would be termed *electronic* ferroelectricity. One important consequence of electronic ferroelectricity would be the lack of piezoelectricity as opposed to normal displacive ferroelectricity where piezoelectricity necessarily appears.

Previous studies on LuFe$_2$O$_4$ have revealed many interesting phenomena such as sequential CO transitions (two dimensional CO at 530 K and three dimensional CO at 340 K) [4,5], a huge coercive magnetic field reaching 10 T at 4 K [6], a large magnetoelectric coupling at room temperature [7], and electric field driven phase transitions [8,9]. It should be noted, however, that most of these studies were concerned with bulk properties as probed by macroscopic measurements, reciprocal mapping by X-ray scattering, neutron scattering, and electron diffraction. In order to further understand the complex physical properties of LuFe$_2$O$_4$, additional microscopic or, in particular, real space measurements would be desired. Detailed investigations of the domain structure of LuFe$_2$O$_4$ would provide new insights for the system. Previously, magnetic domains imaged into the *ab*-plane by magnetic force microscopy (MFM) [6,10] and CO domains viewed from the cross-section plane (containing the *c*-axis) imaged by Transmission Electron Microscopy (TEM) [10-12] were reported. If the origin of the ferroelectricity in LuFe$_2$O$_4$ is indeed CO, the CO domains imaged by TEM can be considered as ferroelectric domains. The three dimensional (3D) structure of the magnetic domains of LuFe$_2$O$_4$ has been proposed based on the *ab*-plane MFM images and the correlation length along the *c*-axis extracted from neutron scattering. The structure of the ferroelectric domains, on the other hand, can only be conjectured due to the lack of the ferroelectric domain structure information viewed from the *c*-direction (*ab*-plane imaging). Once the ferroelectric domain structure in the *ab*-plane is observed, it is possible to properly estimate the 3D ferroelectric domain structure by combining the information with the cross-section CO domain images by TEM. It is also important to compare the ferroelectric domain structure with the *ab*-plane magnetic domain structure to understand possible correlations between the ferroelectric and magnetic orders in LuFe$_2$O$_4$.

In this letter, we report ferroelectric as well as magnetic domain images on the *ab*-plane of LuFe$_2$O$_4$ by various scanning probe microscopy (SPM) methods such as electrostatic force microscopy (EFM) [13], piezoresponse force microscopy (PFM) [14], and MFM [15]. Generally ferroelectric domains can be viewed either with EFM or PFM. PFM utilizes the piezoelectric responses of ferroelectric materials, while EFM probes the surface charges induced by the polarization of the materials. PFM, in particular, is a proper tool to check whether LuFe$_2$O$_4$ is indeed electronic ferroelectricity or not because materials of electronic ferroelectricity are expected to show negligible piezoelectricity. We also present various real space domain images measured by EFM and MFM, and critically compare ferroelectric and magnetic domain structures, both of which turned out to be of irregular shape. We discuss these results with respect to the origin of the irregular domain formation.

**Experimental.** – LuFe$_2$O$_4$ single crystals were grown by a floating zone crystal growing facility (SC1-MDH11020) from Canon Machinery. The typical size of single crystals was 1mm x 1mm x 0.1mm. Magnetic properties and resistivity were measured with a Physical Properties Measurement System (PPMS) from Quantum Design. Ferroelectric and magnetic domains were imaged by EFM and MFM, respectively. PFM measurements were also carried out. Two commercial atomic force microscopes (XE-100, PSIA and UHV-VT SPM, Omicron) were used for EFM, PFM, and MFM measurements. We briefly describe two main non-contact AFM modes used in the present measurements: EFM and MFM.

In the EFM mode, Pt/Ir-coated tips (EFM, Nanoworld) are used as probe [16] and an AC modulation voltage (amplitude $V_{AC}$ and frequency $\omega$) is applied between the tip and the bottom electrode on which a sample is placed. With the modulation voltage in place, the potential difference between the tip and the sample is:

$$\Delta V(t) = V_{AC}\sin\omega t - V_S \qquad (1)$$

where $V_S$ is the surface potential of the sample due to the surface charges and is the information we seek. Then the force between the tip and the sample is given as follows:

$$F(t) = \frac{1}{2}\frac{C}{d}\Delta V(t)^2 = \frac{1}{2}\frac{C}{d}(V_{AC}^2\sin^2\omega t - 2V_{AC}V_S\sin\omega t + V_S^2) \qquad (2)$$

where $C$ is the capacitance of the system and $d$ is the distance between the sample and the tip. From Eq. (2), it is seen that the electric force contains a component modulating at $\omega$ whose amplitude is proportional to $V_S$. Thus, the deflection of the conductive tip appearing in the first harmonic component is proportional to $V_S$, and this deflection is converted to electrical signals through a position sensitive photo diode (PSPD). The electric signals are then filtered by a lock-in amplifier to obtain the first harmonic signal, and EFM images are constructed by collecting the filtered signals at each point.

For MFM measurements, Co-coated AFM tips (MFMR, Nanoworld) are used. Magnetic domain imaging is done via the plane subtraction mode; in this imaging mode, an average topographic plane is first measured and then this topographic plane is used as the offset plane for subsequent magnetic measurements. Magnetic signals (frequency shift $\Delta f$ in the resonant frequency of the cantilever) are obtained by lifting the tip about 20 nm above the offset plane and performing scans. In the present investigation of LuFe$_2$O$_4$, MFM images were collected in this way at various temperatures below the ferrimagnetic transition point. It is noted that the tips used in EFM and MFM measurements were in similar shape and size but coated with different materials. It should be kept in mind, however, that EFM is operated with a tip at constant potential while MFM is performed with a tip at constant charge. This fact will be taken into account when we compare the length scales extracted from EFM and MFM images.

**Results and Discussion.** – The crystal structure of $LuFe_2O_4$ consists of two types of oxide layers; one type is a lutetium oxide layer, and the other one is an iron oxide bilayer. With alternate stacking of these two types of layers the unit cell of $LuFe_2O_4$ is formed as shown in Fig. 1 (a). The iron oxide bilayer plays a predominant role in determining both magnetic and ferroelectric properties of the system. If one focuses on the Fe ions, the bilayer is a double layer of stacked Fe triangular lattices. It is believed that in the ferroelectric phase Fe ions assume two distinct oxidation states, $Fe^{2+}$ and $Fe^{3+}$, and the double layer is divided into $Fe^{2+}$–rich and $Fe^{3+}$–rich layers. The $Fe^{2+}$ ions in the $Fe^{2+}$-rich layer form a honeycomb arrangement with $Fe^{3+}$ located at the center of each hexagon, and vice versa. This particular CO of $Fe^{2+}$ and $Fe^{3+}$ ions then leads to a charge imbalance between the layers within a bilayer and thus produces corresponding electric dipoles and ultimately ferroelectricity in the system [1,3]. Fe ions also carry unpaired spins corresponding to the oxidation states, and the exchange interactions between Fe ions yield ferrimagnetism with geometrical frustration due to the triangular nature of the lattice [17,18]. Fig. 1 (b) shows the magnetization of $LuFe_2O_4$ as a function of temperature. The ferrimagnetic Curie temperature is ~230 K which agrees with previous reports [19,20]. The difference between the out-of-plane and the in-plane moments implies that the magnetic easy axis of this material is parallel to the $c$-axis. This result is in line with the previously proposed model of Ising-type ferrimagnetic order along the $c$-axis [5]. Considering that the magnetization of $LuFe_2O_4$ lies along the $c$-axis, MFM measurements would be performed in the $ab$-plane of the sample. Fig. 1(c) is the resistivity in the $ab$-plane as a function of temperature; the resistivity exhibits an insulating behaviour ($d\rho/dT < 0$) in the whole temperature range. Note that a slope change occurs rather gradually around 340 K as highlighted in the inset. An increase in the absolute value of the slope $|dln\rho/dT|$ in the low temperature phase would be due to the three dimensional CO leading to ferroelectricity [21,22].

Typical ferroelectric domains can be imaged by either EFM or PFM. EFM yields ferroelectric domain structures by detecting the surface electric potential which reflects surface polarization charges. PFM, on the other hand, detects the piezoresponse of ferroelectric domains, a length change in response to an applied voltage occurring in ferroelectric materials. Electronic ferroelectricity, however, would not respond piezoelectrically because electronic ferroelectricity is due to electronic charge redistribution in the unit cell. We attempted to image the ferroelectric domains of $LuFe_2O_4$ by both techniques. It is known that the electric polarization of $LuFe_2O_4$ is also parallel to the $c$-axis [1]. Therefore, SPM tips should be brought to the $ab$-plane of $LuFe_2O_4$ to probe the responses. It is technically critical to secure clean flat surfaces for SPM measurements; the layered structure of $LuFe_2O_4$ allows cleaving and thus flat clean $ab$-plane surfaces are easily obtained. Cleaved $LuFe_2O_4$ surfaces were first scanned by both EFM and PFM at room temperature. The ferroelectric domain structures were indeed observed by EFM and a typical image is shown in Fig. 2 (a). The red and blue regions in the figure correspond to the surface charges due to up and down polarizations, respectively. The ferroelectric domains have irregular patterns, and the average size of the domains is less than 100 nm. This unusual domain structure is discussed below in detail. For now let us move on to the PFM results; Fig. 2 (b) is the image of the same surface scanned by PFM measuring the vertical polarization component. It is remarkable that vertical PFM does not reveal any sign of a domain structure at all. Various efforts to improve the measurement sensitivity by changing frequency and amplitude of the AC probing voltage yielded the same image. This result obviously means that $LuFe_2O_4$ lacks piezoelectricity. Thus, the ferroelectricity of $LuFe_2O_4$ is without piezoelectricity, but still generates surface charges. The observation of ferroelectricity without any piezoresponse in turn implies that the ferroelectricity in $LuFe_2O_4$ is not of usual displacive type and suggests that its origin is indeed electronic.

The ferroelectric/charge order transition in $LuFe_2O_4$ has previously been studied by various methods. The pyroelectric current measurement [1] and the temperature dependence of charge order superlattice spots [5] are representative examples in determining the phase transition temperature. In order to confirm the same transition by the SPM technique, EFM measurements were conducted as a function of temperature in heating direction from room temperature. Fig. 3 (a) and (b) are the EFM images taken at 314 K and 390 K, respectively. In comparing the two images, the first thing to notice is a large difference in the contrast of the signal strength represented by the color images. This difference is attributed to the disappearing polarization as temperature increases above $T_{FE}$. According to the TEM study [5], there still exists two dimensional CO even above $T_{FE}$. In order to obtain further information, EFM measurements at several temperatures are compared. Fig. 3 (c) is a display of the average contrast as a function of temperature. The average contrast is obtained from a standard deviation in the signal strength histogram of each EFM image. It is seen that the contrast suddenly decreases between 330 K and 350 K. According to the resistivity data in Fig. 1 (c), the ferroelectric CO transition temperature of $LuFe_2O_4$ is located around 340 K. Thus, we may conclude that the sudden drop in EFM image contrast is caused by the phase transition.

Now returning to the irregularity of the ferroelectric domains and its origin, it is noted that $LuFe_2O_4$ is known to have several kinds of defects such as non-stoichiometric defects [23] and stacking faults [5]. Thus, it is important to check whether these defects are responsible for domain formation in any way. For this purpose, we have imaged by EFM exactly the same surface area at room temperature before and after annealing above the transition temperature. Fig. 4 (a) and (b) are the domain images before and after annealing, respectively. It is easily seen from the figures that the general shape and qualitative features of the domain patterns remain unchanged but the exact locations of up and down domains have changed. If the domain formation is initiated by the defects in the sample and the domains are pinned by them, the locations of up and down domains in the two domain patterns should remain the same or at least similar because the annealing temperature is not high enough to change the defect positions. Thus, noting that the qualitative nature of the two domain patterns before and after annealing remain the same despite the fact that the exact locations of up and down domains are not reproducible, we conclude that the origin of the domain formation in $LuFe_2O_4$ is intrinsic rather than extrinsic. The intrinsic property such as geometrical frustration inherent in the crystalline structure would be responsible for the irregular domain formation.

Having identified the general features of the ferroelectric domains, we wish to compare the ferroelectric and magnetic domains of LuFe$_2$O$_4$. In Fig. 5(a) and (b) displayed are the ferroelectric domains obtained by EFM and the magnetic domains by MFM, respectively. Fig. 5(a) and (b) were obtained by the same AFM at approximately 100 K, and the size of the images are the same as indicated by the scale bars in the figures. Note that the EFM signal strength represented in Fig. 5(a) is different from those of the previous EFM images because a different AFM and a different sample were used in the measurements. Nevertheless, the meaning of the colors remain unchanged. The MFM signals, represented in Fig. 5(b), were obtained as a frequency shift, and here again the red and blue colors denote the up and down magnetic domains, respectively. Although the domain structures in the ferroelectric and magnetic images are similar in shape, their average domain sizes are rather different. The similarity of the domains in shape would suggest the common cause. As mentioned in the introduction, both properties result from the ordering of Fe$^{2+}$ and Fe$^{3+}$ ions. For the magnetic domains, W. Wu *et al.* [5] proposed a model of pancake-like domains with disorder as the reason for the irregular magnetic domain pattern. We may also apply the pancake-like domain model to the ferroelectric domains. It should still be kept in mind, however, that the intrinsic geometrical frustration is indispensible for the irregular ferroelectric domains. The most distinctive difference between the ferroelectric and magnetic domains is their average size. The mean length of the irregular ferroelectric domains is estimated to be ~100 nm; on the other hand, the size of the magnetic domains varies widely and the linear dimension even reaches ~1 μm. This large difference in length scale between the ferroelectric and magnetic domains is striking and visually conspicuous in Fig. 5(a) and (b).

At this point, we may deal with a resolution issue. EFM and MFM are performed in different operating modes with respect to the probing tip configuration: constant potential vs constant charge. We have simulated and compared the resolution of the two modes utilizing the commercial COMSOL Multiphysics software with real tip parameters as input [24]. For the EFM and MFM tips from Nanoworld with radius of curvature 20 ~ 40 nm [16], it turns out that EFM in constant potential mode and MFM in constant charge mode show just about the same resolution capability. Thus, the domain size difference is not due to the resolution difference in the two operating methods, and should be regarded as a real difference in length scale.

Recognizing the size difference in the ferroelectric and magnetic domains of LuFe$_2$O$_4$, we attempt to explain the domain size difference in the following way. As explained above, in LuFe$_2$O$_4$ the ferroelectric polarization is developed by separating an iron oxide bilayer into Fe$^{2+}$-rich and Fe$^{3+}$-rich layers, while the magnetization is determined by the direction of the magnetic moments of Fe$^{2+}$ ($S$ = 2) and Fe$^{3+}$ ($S$ = 5/2) which are not cancelled out in the ferrimagnetic configuration. Consider, as shown in Fig. 5(c), a bilayer which has two ferroelectric domains with opposite polarization. According to Ref. [18], the spins in the iron oxide bilayer of LuFe$_2$O$_4$ in the magnetic phase is arranged in such a way that Fe$^{2+}$ spins in both Fe$^{2+}$-rich and Fe$^{3+}$-rich layers are all parallel while Fe$^{3+}$ spins in the Fe$^{2+}$-rich layer is antiparallel to the Fe$^{2+}$ spin direction and Fe$^{3+}$ spins in the Fe$^{3+}$-rich layer are antiparallel to each other. Consequently, the magnetic moments in the Fe$^{2+}$-rich and Fe$^{3+}$-rich layers point in the same direction as depicted in the figure. Note that the magnetic moment in the Fe$^{3+}$-rich layer is larger than that in the Fe$^{2+}$-rich layer. In this situation, EFM would be able to resolve the two domains while MFM would not. MFM detects magnetic force arsing between the tip and the sample, and the magnetic force on the tip resulting from the left and right domains in Fig. 5(c) are indistinguishable. Generally an appearance of magnetic domain walls inside a ferroelectric domain would be excluded considering the high domain wall energy of $LuFe_2O_4$ due to its exceptionally high magnetocrystalline anisotropy [6].

**Conclusions.** – In conclusion, we have imaged the ferroelectric and magnetic domain structures within the *ab*-plane of $LuFe_2O_4$ by SPM techniques which are non-destructive measurement tools. The simultaneous PFM and EFM measurements have revealed the non-displacive, electronic ferroelectricity in $LuFe_2O_4$. EFM measurements as a function of temperature have identified a phase transition around 340 K which agrees with the previous results measured by other means. The annealing effects on the domain formation were also investigated. The location of up and down ferroelectric domains changed completely through the annealing process while the qualitative shape and features remain the same. This result indicates that defects are not the main player for the domain location in $LuFe_2O_4$. Comparison of the ferroelectric and magnetic domain patterns has revealed a fact that the domain patterns share a similarity in shape but a difference in size exists. The former may be attributed to the geometrical frustration and charge disorder affecting the ordering of Fe ions which determines the electric and magnetic properties of $LuFe_2O_4$. The latter could result intrinsically from the different fluctuation behaviors of charge and magnetic degrees of freedom.

***

YHJ wishes to thank S. Ishihara for useful discussions. This work was supported by the National Research Foundation (Grant No. 2011-0009231) and the Center for Topological Matter at POSTECH (Grant No. 2011-0030786). SBK and SWC are supported by the Max Planck POSTECH/KOREA Research Initiative Program (Grant No. 2011-0031558) through NRF. SWC is also supported by the NSF under Grant No. NSF-DMR-1104484.

Figure Captions

Fig. 1: (Colour on-line) (a) Crystal structure of $LuFe_2O_4$. (b) Temperature dependence of the magnetization parallel and perpendicular to the c-axis. Both field cooled (FC) and zero field cooled (ZFC) measurements with applied field 1 kOe were performed. (c) Temperature dependence of in-plane resistivity. A slope change, as shown in the inset, occurs around 340 K due to a ferroelectric charge ordering tranistion. The absolute slope was obtained from local fitting as indicated by solid lines.

Fig. 2: (Colour on-line) (a) EFM image and (b) vertical PFM image of a cleaved *ab*-sruface of $LuFe_2O_4$ at room temperature. Red and blue colors in the EFM image indicate up and down domains of polarization, respectively.

Fig. 3: (Colour on-line) EFM images obtained at (a) 314K and (b) 390K. The image size is ~800 × 800 $nm^2$. (c) Contrast, defined as a standard deviation in the signal strength histogram, is shown as a function of temperature. There is a break between 330 and 350 K. The solid lines are a guide to the eye.

Fig. 4: (Colour on-line) EFM images at room temperature (a) before and (b) after annealing above the transition temperature 340 K on a cleaved surface of $LuFe_2O_4$.

Fig. 5: (Colour on-line) (a) EFM and (b) MFM images of a $LuFe_2O_4$ surface at 100 K. The large difference in length scale between the ferroelectric and magnetic domains is striking. Red and blue colors in the images indicate up and down domains of polarization or magnetization. (c) A schematic diagram of an iron oxide bilayer for a situation where a ferroelectric domain wall forms within a magnetic domain.

Figures

Figure 1

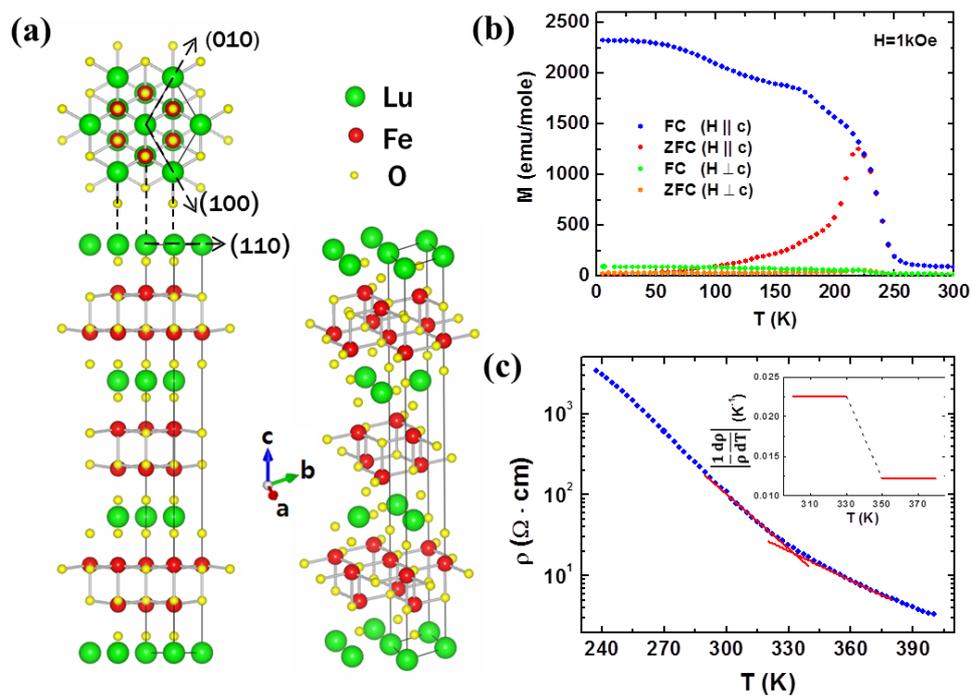

Figure 2

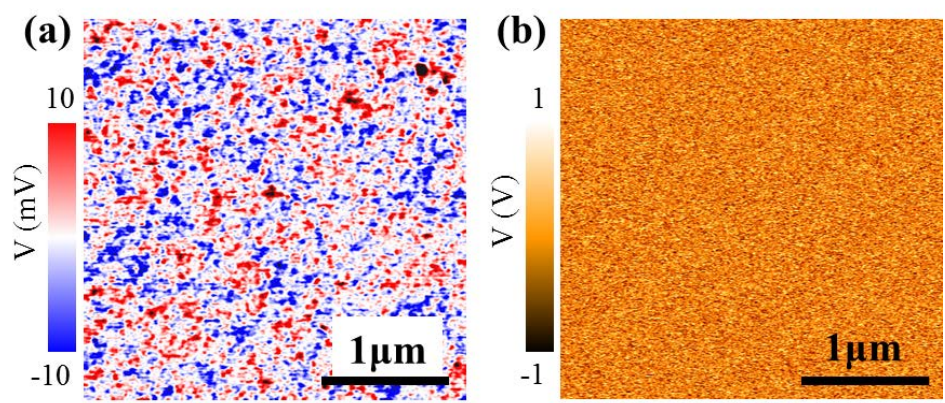

Figure 3

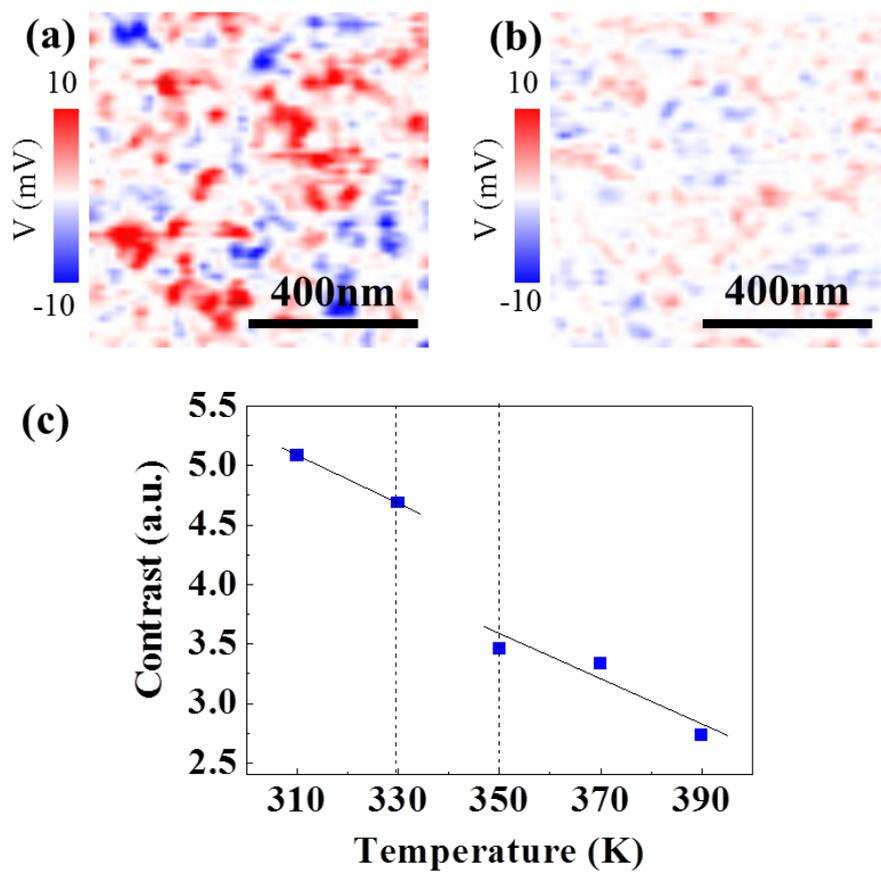

Figure 4

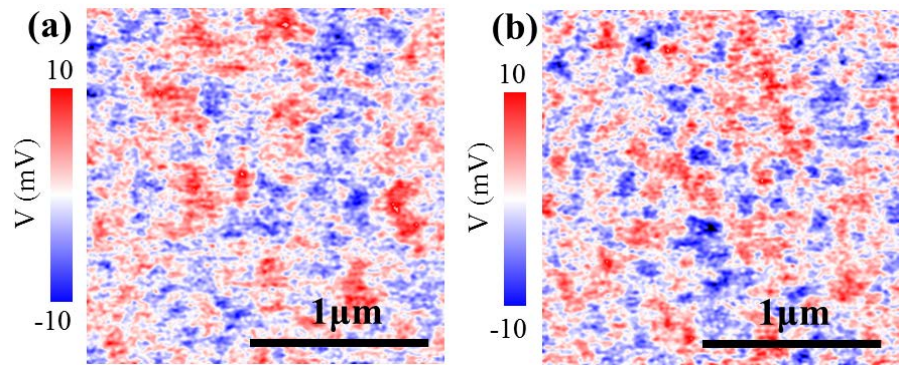

Figure 5

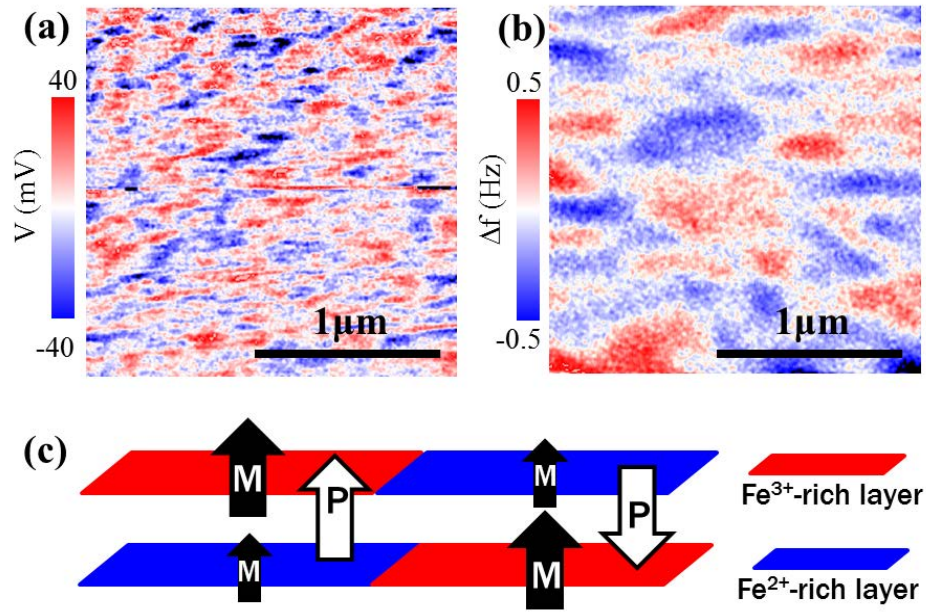